\documentclass{raa}
\usepackage{graphicx,times}
\usepackage{natbib}
\usepackage{amssymb,amsmath}

\bibpunct{(}{)}{;}{a}{}{,}

\usepackage[a4paper=true,dvipdfm=true,pagebackref=true]{hyperref}
\hypersetup{pdftitle = The title of my PDF, pdfauthor = My name, pdfsubject= The subject, pdfkeywords = keyword1 keyword2 keyword3}
\hypersetup{colorlinks = true, linkcolor = green, anchorcolor = red, citecolor = blue, filecolor = red, pagecolor = red, urlcolor = red}

\begin{document}

\title{The Co-alignment of Winged H$\alpha$ Data Observed by the \emph{New Vacuum Solar Telescope}
\,$^*$
\footnotetext{$*$ Supported by the National Natural Science Foundation of China. }
\footnotetext{$\dagger$ Corresponding author}
}

 \volnopage{ {\bf 20XX} Vol.\ {\bf X} No. {\bf XX}, 000--000}
   \setcounter{page}{1}


   \author{Yun-Fang Cai\inst{1}, Xu Yang\inst{2}, Yong-Yuang Xiang\inst{1}, Xiao-Li Yan\inst{1}, Zhen-Yu Jin\inst{1}, Hui Liu\inst{1}, Kai-Fan Ji\inst{1}\,$^\dagger$    }

\institute{Yunnan Astronomical Observatory, National Astronomical Observatories, Chinese Academy of Sciences,
           Kunming 650011, China; {\it jkf@ynao.ac.cn}\\
          \and
           Big Bear Solar Observatory, New Jersey Institute of Technology, 40386 North Shore Lane, Big Bear City, CA, USA 92314\\
\vs \no
   {\small Received 20XX Month Day; accepted 20XX Month Day}
}

\abstract{The \emph{New Vacuum Solar Telescope} (NVST) has been releasing its novel winged H$\alpha$ data (WHD) since April 2021, namely the H$\alpha$ imaging spectroscopic data. Compared with the prior released version, the new data are further co-aligned among the off-band images and packaged into a standard solar physics community format. In this study, we illustrate the alignment algorithm used by the novel WHD, which is mainly based on the optical flow method to obtain the translation offset between the winged images. To quantitatively evaluate the alignment results of two images with different similarities, we calculate the alignment accuracies between the images of different off-band and line center, respectively. The result shows that our alignment algorithm could reach up to the accuracy of about 0.1\,$''$ when the off-band of winged image is lower than 0.6\,\AA\,. In addition, we introduce the final product of the WHD in detail, which can provide convenience for the solar physicists to use high-resolution H$\alpha$ imaging spectroscopic data of NVST.
}

   \authorrunning{Y.-F. Cai et al. }             
   \titlerunning{The Co-alignment of Winged H$\alpha$ Data}  
   \maketitle

%
\section{Introduction}           
\label{sect:intro}
The two-dimensional (2D) imaging spectroscopy observations always play an important role in solar physics studies, which can obtain not only the multi-wavelength spatial images but also the spectral information of the observation target (\citealp{Huang1995}).
There are two classical methods to acquire 2D imaging spectroscopy: one is based on spatial scanning with a slit spectrograph to take a series of spectrograms (\citealt{2003Hanaoka}, \citealt{2003Hanaoka}, \citealt{Chae2013}, \citealt{Li2019}). The advantage of this method is its high spectral resolution and free spectral range, which is capable of investigating the high-speed and multi-atmospheric layer solar features. The 2D spectroscopic images of different wavelength, which are subsequently synthesized by the scanning spectrograms, are strict matching within a single scan in principle, because each spectrum is the performance of a same position on the solar surface within the wavelength range of a spectrogram(\citealt{2018RAAcai}). The other method takes monochrome images in different wavelengths based on the narrow-band filtergrams (\citealp{Bonaccini1990}, \citealp{VTT2002}, \citealt{Liu2014}, \citealp{Deng2019}). It directly delivers high spatial resolution monochromatic images with different off-band. However, considering the temporal resolution cannot be too low, there only several off-bands are selected for observation generally. Because of the influence of the near-ground atmospheric turbulence and telescope mechanics, the monochromatic images of off-band observed by this method are nonalignment usually for ground-based solar telescopes, which requires subsequent processing of alignment.

\begin{figure}
\centering
\includegraphics[width=15cm]{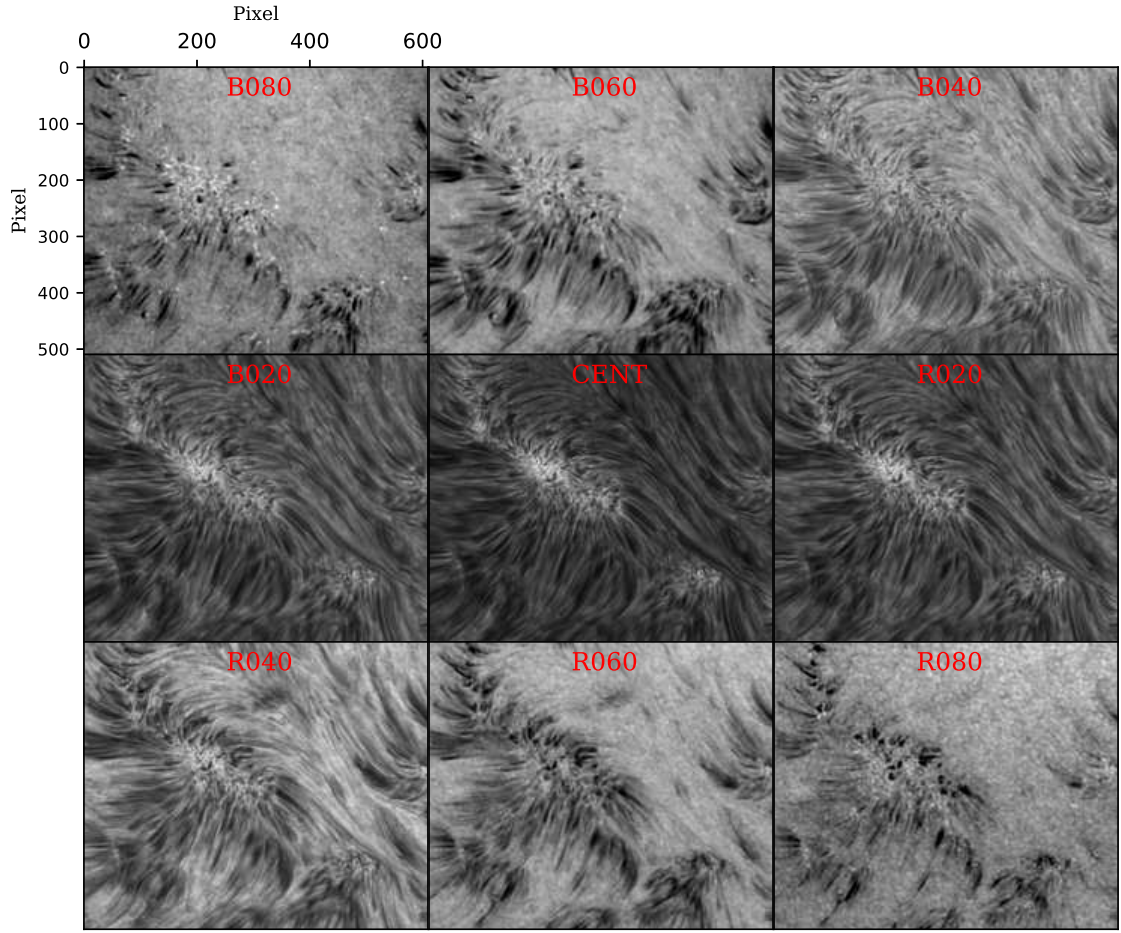}
\caption{The local high resolution monochromatic images of 9 off-bands (B080 corresponds to the blue wing 0.8\,{\AA} and R080 corresponds to the red wing 0.8\,{\AA} and so on).}
\label{Fig1}
\end{figure}

The H$\alpha$ channel imaging system of NVST is based on the mode of filtergrams, which can realize the off-band observation in the range of $\pm$ 4\,{\AA} by a tunable Lyot filter with a bandwidth of 0.25\,\AA.
To achieve a high temporal resolution, the most off-band observations of H$\alpha$ is the three bands mode (i.e., two symmetrical off-bands and one line center). The off-band wavelength is usually selected in the range of 0.3\,--\,0.6\,{\AA} during observation.
There always exist a translation transformation between different off-band monochromatic images, and this translation offset can sometime reach 5-10 pixels (0.8\,--\,2\,$''$).
The image scale of the H$\alpha$ winged images is 0.164\,$''$, so the alignment accuracy is always required to reach sub-pixel level of about 0.1 arcsec.
The old version of WHD released only did the preprocessing and high-resolution image reconstruction, and it was no co-alignment among the off-band images. For further scientific research, the winged images whose wavelength close to the line center are aligned roughly by the cross-correlation alignment, which method always failed for the far winged data because of the low similarity between the images of winged and line center.
In the Figure~\ref{Fig1}, we show the local high resolution monochromatic images of 9 different off-bands. It can be seen that the greater wavelength of off-band, the lower similarity between the images of winged and line center.
Therefore, we urgently need to study an effective algorithm to high-precision align the off-band images. For the three-band observations, we should calculate the alignment accuracy of different off-band image relative to line center one, and find out which off-bands can meet the accuracy requirements of sub-pixel. All those works have important guiding significance for the NVST routine off-band observations.

The multi-band observation and research of the solar atmosphere fine structure and the small-scale magnetic field have become an important topics in solar physics, which requires a high alignment accuracy among multi-band images. There are two registration methods of the solar multi-band images used commonly: one is based on cross-correlation alignment (\citealp{Kuehner2010}) using the similar feature structure between images, and the alignment accuracy (about 0.1\,--\,0.5\,$''$) (\citealp{Berkebile-Stoiser2009}) is severely limited by the image structure.
For the images with large difference, the other alignment of feature point matching is the most commonly used (\citealp{Guglielmino2010}). In this method, the more feature points, the higher the alignment accuracy. However, the feature points of off-band image are artificially selected with a certain physical model at present, so there are often too few feature points and too much manual intervention to improve accuracy. The alignment accuracy of this method is about 1 - 2\,$''$ and completely depends on the characteristics of the images (\citealp{2006Smith}).
In this paper, we propose a new co-alignment method of winged images based on the optical flow, which is very popular algorithm for motion image analysis. The method has higher accuracy and lower dependence on similar structures compared with cross-correlation alignment, which have successfully applied to the alignment among high-resolution monochromatic images of the NVST H$\alpha$ channel.

The paper is organised as follows: In Section ~\ref{S:1}, the instrumentations and observations of winged images are described, and then the alignment method of optical flow is exhibited in Section ~\ref{S:2}. In Section ~\ref{S:3}, we asses the alignment accuracies of our algorithm for the different off-band image relative to the line center one. The novel WHD product is introduced in Section ~\ref{S:4} and the conclusion and discussion are given in Section ~\ref{S:5}.


\section{Instrumentations and Observations}
\label{S:1}
The NVST is a vacuum solar telescope equipped with a 985 mm clear aperture primary mirror, operating by Fuxian Solar Observatory (FSO)/ Yunnan Observatories (\citealt{Liu2014}). Its scientific goal is to obtain the solar information by the high resolution imaging (\citealp{Xiang2016}) and spectral observations (\citealt{Xu2014}, \citealt{Cai2017}) in the wavelength ranges from 0.3 to 2.5\,$\mu$m. There are two terminal instruments: the Multi-band High Resolution Imaging System (MHRIS) and the grating spectrometer system. The MHRIS includes two broad-band interference filter channels (TiO-band and G-band) and three narrow-band Lyot filter channels (H$\alpha$, Ca II 3933\,\AA\ and He I 10830\,\AA). The grating spectrometer system (\citealt{2018RAAcai},\citealt{Cai2018principal}) consists of the Multi-band Spectrograph (MBS) in visible lines and the High Dispersion Spectrometer (HDS) in near-infrared lines.
In this paper, we focus on aligning the H$\alpha$ winged data of MHRIS.

The NVST is a ground-based solar telescope, and the adaptive optics (AO) systemm (\citealt{Rao2016}) are not yet be widely used in daily observations. In order to overcome the influence of seeing, the speckle masking technique (\citealp{Liu1998},\citealp{Xiang2016}) is used to obtain the high resolution monochromatic images of each off-band.
Here, we take a set three off-band images as an example for alignment. The data were observed on 26 February 2021 over a quiet sun region. The offset wavelength are $\pm$\,0.4\,\AA\ and H$\alpha$ line center, respectively. The size of image is 1024 $\times$ 1024 pixels$^{2}$ and the scale of each pixel is 0.164\,$''$, so the image covers a field of view (FOV) of 168\,$''$ $\times$ 168\,$''$. A high resolution image of each off-band is reconstructed by 100 frames short exposure images which is a single frame image taken at an exposure time of 20 ms. The spatial resolution of a high resolution image can reach about 0.1\,--\,0.3\,$''$. The cadence of one winged and H$\alpha$ line center images is about 14 s, and it takes about 40 s in total for a single off-band scanning in the routine mode of WHD.

\section{Alignment Method of the Winged Images}
 \label{S:2}
The alignment method proposed in this paper includes three processes: First, we apply the optical flow method to calculate the relative motion field between the imaging spectroscopic dataset. Then, to improve the alignment accuracy, the method of histogram statistics is used to obtain the system translation between the winged images. Furthermore, we introduce the weighted least-squares to achieve a final high-precise offset. The specific methods and steps are as follows.

\subsection{Calculate the Relative Movement Field}
 \label{S:2:1}
\begin{figure}
\centering
\includegraphics[width=15cm]{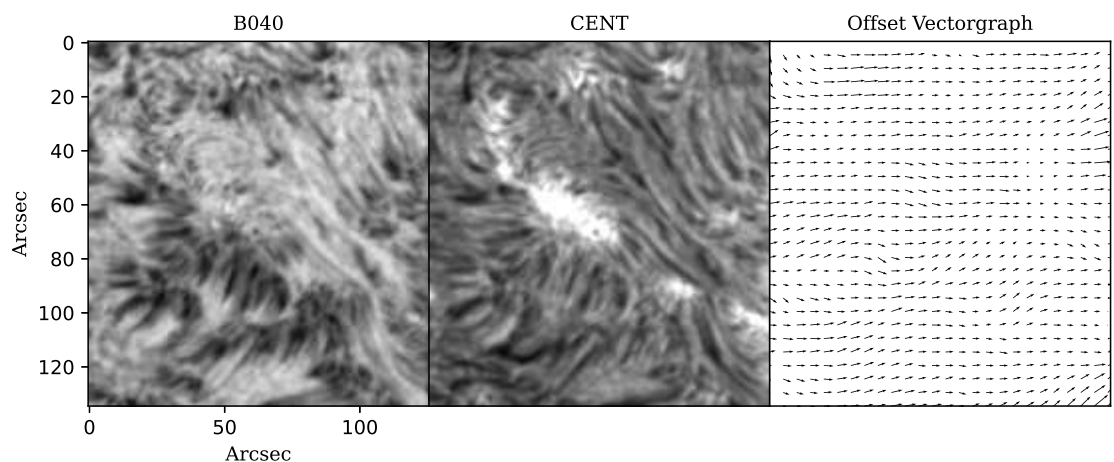}
\caption{The local monochromatic images of H$\alpha$ blue wing 0.4\,\AA\,, line center and their offset vectorgraph (B040 relative to line center), 15 pixel samples is used to display the offset distribution conveniently.}
\label{Fig2}
\end{figure}

The optical flow, as an important research direction of computer vision, describes the instantaneous apparent motion of the object (\citealt{Horn1981}). This method is mainly used in object recognition and object tracking to obtain the object motion information from the corresponding relationship between two neighboring image frames in the time domain. Compared with general natural images, the observed solar image is more complex, which always contains many different activity phenomena. The motion of these phenomena are not only nonrigid and random, but also often have a large range. There are many kinds of theoretical bases and mathematical methods for calculating optical flow. The Gunnar farneback's algorithm (\citealt{Farneb2003}) is a kind of dense optical flow, which can calculate the motion information of pixels one by one and generate a Gaussian Pyramid of images with different resolutions to perform the multi-resolution image search. Although this algorithm is more time-consuming than other optical flow methods (such as the  sparse optical flow), it can obtain a higher calculation accuracy for the images with complex structure and motion characteristics, so we chose it to calculate the motion information of each pixel for the solar images.
The Gunnar farneback's algorithm is complex in calculation and requires huge computation power. However, with the development of computer application ability in recent years, many computer languages have encapsulated the algorithm into a standard code base. In this work, we apply the standard OpenCV calcOpticalFlowFarneback processing package of the python to realize the operation above.
The left and middle panels of Figure~\ref{Fig2} present the local monochromatic images of H$\alpha$ blue wing 0.4\,\AA\, and line center, respectively, and the right panel shows their relative movement field (offset vectorgraph) generated by the above algorithm of optical flow. From the figure we can clearly see the motion information of pixel.

\subsection{Obtain the System Translation of Winged Images}
 \label{S:2:2}
The images of H$\alpha$ winged and line center are observed in a coaxial focal plane system. Therefore, the distortion between them is primarily caused by the translation transformation.
Theoretically, the movement field calculated by the optical flow reflects all the motion of each pixel, including not only the system translation of the image, but also the evolution of solar structure.
However, because the time delay of the adjacent off-band and line center image is too short for major solar feature evolution, that is the evolution of solar structure only occurs in a small number of pixels, but the system translation of image exists in all the pixels.
Therefore, we can retrieve the system translation of the image by the statistical method.

\begin{figure}
\centering
\includegraphics[scale=0.7, angle=0]{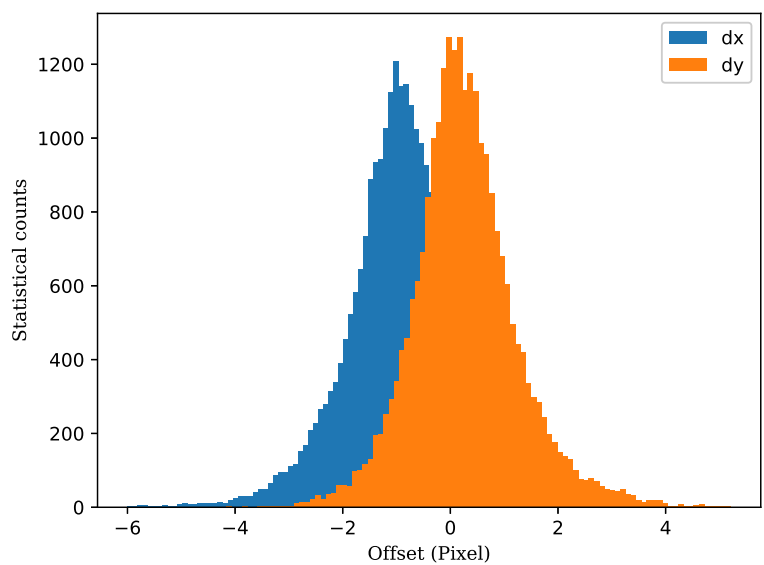}
\caption{The histogram statistics for the movement field from Figure~\ref{Fig2} right panel in X and Y directions, respectively.}
\label{Fig3}
\end{figure}

Figure~\ref{Fig3} presents histogram statistics for the movement field from Figure~\ref{Fig2} right panel, in X and Y directions, respectively. To reduce random errors, we use the following steps to obtain the system translation of winged images: Firstly, we calculate the position of the center of gravity within a fixed range near the peak of histograms as the reference position. Then, we define the points within five pixels from the reference position as the effective points and discard points out of this range. It is necessary to ensure that the pixel is an effective point in both the X and Y directions. Finally, we measure the median value of all effective points as the system translation in X and Y directions, respectively.

 \subsection{Achieve the Final Offset of Winged Images}
 \label{S:2:3}
As is well known, the higher the similarity of the two images, the higher the calculation accuracy of their movement field (\citealp{2006Smith}), and the more effective points in the fixed range. Therefore, we can make full use of the number of effective points to achieve a high-precision final offset of winged images. That is, we can get three groups system translations between the three off-band images with the above method (B040 to center, R040 to center, and B040 to R040), and make the system translation which has more effective points to play a dominant role in determining the final offset. To achieve this purpose, solving the weighted least-squares is an excellent way.

Besides the system translations obtained with the above method, we also calculate the corresponding effective point proportions between two of the three spectroscopy images, respectively. The effective point proportion is the ratio of the number of effective points to the total number of pixels. Then, take the effective point proportions as the weighting coefficient of the least-squares method to obtain the final offsets of each off-band relative to each other. Table~\ref{Tab0} lists the proportion of effective points, system translation and final offset between the images of H$\alpha$ blue/red wings and line center. The two values in the square brackets of the system translation and final offset represent the offset in the X and Y axes, respectively. It shows that there is a higher proportion of effective points between the winged and line center image (0.891 and 0.856) than between two winged images (0.6). The system translation and the final offset are close but not identical, which plays a certain role in improving the alignment accuracy.

\begin{table}
\bc
\caption[]{The proportion of effective points, system translation and final offset between the images of H$\alpha$ blue/red wings and line center.}
\label{Tab0}


 \begin{tabular}{c|c|c|c}
   \hline
        & Proportion  & System Translation  & Final Offset   \\
         &   & (pixels)  & (pixels)    \\
  \hline
B040-CENT   & 0.891  & [ -1.56, -1.878 ]  & [ -1.532, -1.914 ]   \\

  \hline
R040-CENT  & 0.856  & [ 1.402, 2.002] & [ 1.372, 2.042 ]  \\

  \hline
B040-R040 & 0.6  & [ -2.843, -4.037 ]  & [ -2.904, -3.956 ]  \\

  \hline
\end{tabular}
\end{center}
\end{table}

\section{Alignment Accuracy}
 \label{S:3}

The asynchronous datasets from filter-based imaging spectroscopy are insufficient to assess the algorithm's alignment accuracy. However, taking advantage of the grating-based spectrograph with a slit, we can create a test dataset with its synchronized 2D spectroscopic images and customized offsets. In this way, the alignment accuracy can be evaluated by the pre-set translation displacements and the measured final offsets. In this research, we applied our algorithm to the 2D spectroscopic images observed by the Fast Imaging Solar Spectrograph (FISS, \citealp{Chae2013}) onboard the Goode Solar Telescope (GST,\citealp{Cao2010}). To verify the universality of the algorithm, we select the H$\alpha$ spectrum over a quiet Sun region (the alignment of quiet region is more difficult than that of region with obvious solar activities).
The FISS observation was taken on June 8, 2021. The spectral resolution near the H$\alpha$ line is 0.02\,\AA \,pixel\textsuperscript{-1}, and the wavelength range is about 10\,\AA. The spatial resolution of FISS is 0.16\,$''$ in the slit direction and in the scan direction.

\begin{figure}
\centering
\includegraphics[width=15cm, angle=0]{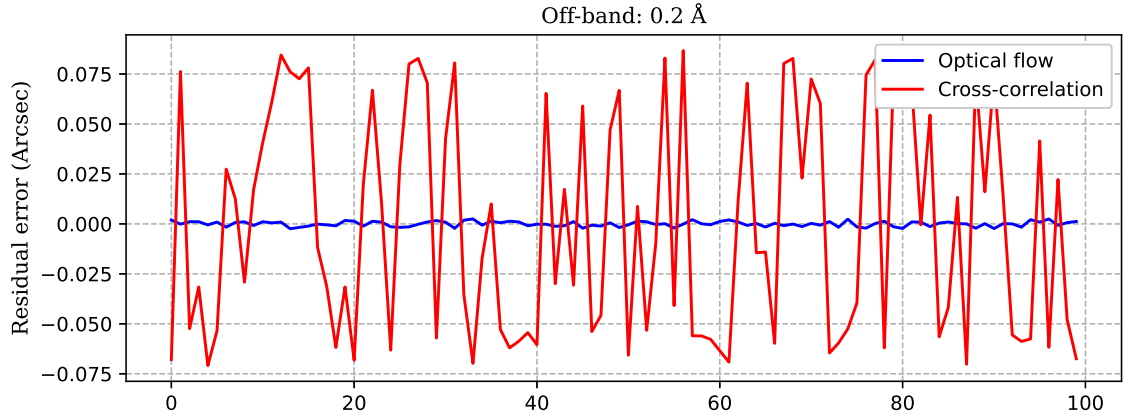}
\caption{The residual error of the calculated and the known translation offset of H$\alpha$ blue wing 0.2\,\AA\, relative to line center, the red and blue lines represent the results of cross-correlation and optical flow, respectively.}
\label{Fig5}
\end{figure}

The specific steps to evaluate the alignment accuracy are as follows: Firstly, the two FISS images of winged and line center are composed with the bandwidth of 0.25\,\AA\, (same as NVST) as the original dataset. Second, to reduce the randomness of the results, we randomly generate 100 groups of 2D floating-point numbers (x and y axis) as the known translation offsets, and shift the winged image 100 times relative to line center one with the known translation offsets in x and y axis respectively, to obtain a test set. Then, two alignment methods of the optical flow and the cross-correlation are used to calculate translation offsets for comparison with each other. In Figure~\ref{Fig5}, we show the 100 groups residual error of two alignment methods in the X-axis of the off-band 0.2\,\AA\, image. This residual error is the difference between the measured and known translation offset. From the figure, we can clearly see that the residual error of optical flow method is much smaller than that of the cross-correlation.
%
Finally, the root-mean-square error(RMSE) of 100 groups known and measured translation offset are calculated as the alignment accuracy of the two images.
According to the above steps, we compose 10 winged images with different off-band and evaluate their alignment accuracy relative to the image of line center, respectively.
As the alignment accuracy is closely related to the image structures, we also calculate the similarity which is expressed by the correlation coefficient of two images with the maximum value of 1. The larger the value, the higher the similarity between two images.
In the Table~\ref{Tab1}, we list the similarity and corresponding RMSE of 10 off-bands, the $\rm RMSE_{\rm opt}$ and $\rm RMSE_{\rm cc}$ represent the accuracies of the optical flow and cross-correlation method, respectively.
Here, we only list the results of H$\alpha$ blue wings, because the measured results of red and blue symmetric off-bands relative to the line center are very similar.
\begin{table}
\bc
\caption[]{ The Similarity and Alignment Accuracy of 10 H$\alpha$ Off-band Images Relative to Line Center.}
\label{Tab1}


 \begin{tabular}{c|ccccccccccc}
  \hline
Off-band    & -1.0  & -0.9  &-0.8  &-0.7  &-0.6  &-0.5  &-0.4  &-0.3  &-0.2  &-0.1  &0  \\
(\AA)& &&&&&&&&&&\\
  \hline
Similarity  &0.270  & 0.297 & 0.269 &0.208 &0.217 &0.305	&0.422 &0.581 &0.788 &0.920	&1.0 \\
       & &&&&&&&&&&\\
  \hline
$\rm RMSE_{\rm opt}$&0.780  &0.862  &1.084  &0.627  &0.269  &0.102 &0.095 &0.016 &0.004 &0.002 &0    \\
(arcsec)& &&&&&&&&&&\\
  \hline
$\rm RMSE_{\rm cc}$ &6.458  &6.339  &6.983  &6.147  &6.039  &2.253  &0.871 &0.340 &0.068 &0.053 &0    \\
(arcsec)& &&&&&&&&&&\\
  \hline
\end{tabular}
\end{center}
\end{table}
It can be seen from Table~\ref{Tab1} that the image similarity is larger and the RMSE of both methods are all relatively small at the wavelength near from the line center. For example, the similarity is about 0.788 at the off-band of 0.2\,\AA\,, and the alignment accuracy of optical flow and cross-correlation can reach up to 0.004\,$''$ and 0.068\,$''$ respectively. However, the similarity decreases with the increase of off-band, and the accuracy of both methods gradually decreases too, especially the cross-correlation method.
As for the off-band 0.6\,\AA\,, the winged image has a low similarity (0.217) with the line center one.
The alignment error of the cross-correlation is about 6\,$''$ but it still can reach about 0.3\,$''$ with the optical flow algorithm, which indicates the cross-correlation method has failed at this off-band, and the dependence of optical flow alignment on the image similarity is not as strong as that of cross-correlation method.
In general, our optical flow algorithm works more accurately than the cross-correlation method. It could reach up to the accuracy of within 0.1\,$''$ when the similarity is more than 0.3 which corresponds to the maximum off-band is about 0.5\,\AA\,. If the off-band is more than this value, the alignment of optical flow does not meet the accuracy requirements of sub-pixel, so it is necessary to increase some intermediate off-band images as reference images for transition alignment.
It is worth mentioning that the original data set used for accuracy calculation are real observations in the solar quiet region, which barely contains obvious solar structures. With stronger solar feature (sunspot or filament), we can expect a higher off-band of about 0.6\,\AA\,or higher with the alignment accuracy of 1 pixel.

\section{The Products of NVST Winged H$\alpha$ Data}
 \label{S:4}

The winged H$\alpha$ data of NVST is processed with the flat-field, dark correction, high resolution reconstruction with the speckle masking technique. Alignment between off-band images with the method of optical flow described in the paper is now applied as well.
In addition, the sequence images of single band are also be aligned. As the image structures of two adjacent frames in the same wavelength sequence are similar, their alignment can be well achieved by either the cross-correlation or the optical flow method.
To ensure higher accuracy, we also use the optical flow alignment for sequence images.
After the above processing, the well-aligned and scientifically ready winged H$\alpha$ images are packaged in three standard forms: datacube, picture, and movie.

\begin{figure}
\centering
\includegraphics[width=13cm, angle=0]{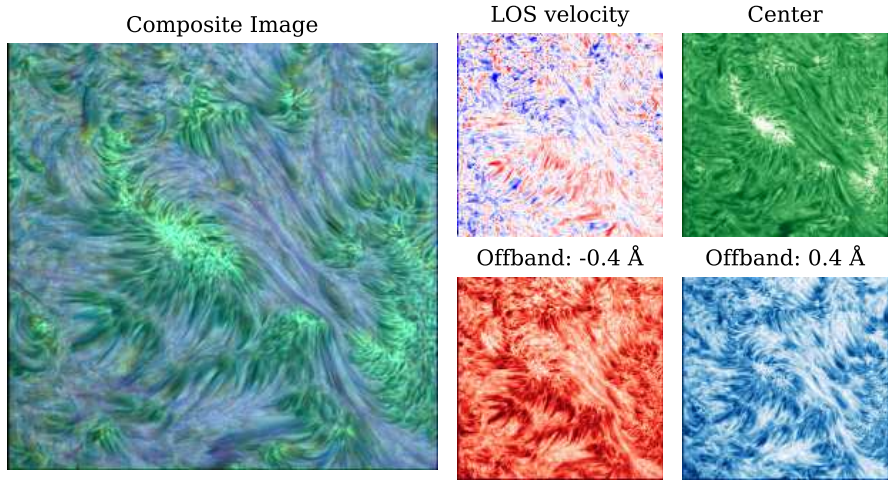}
\caption{The winged H$\alpha$ data of the NVST.}
\label{Fig7}
\end{figure}

The scientific data is a FITS file of three-dimensional datacube, and each FITS is composed with pseudo-color off-band images. The detailed information of observation data can be obtained from a standard FITS header, including observation time, image size, exposure time, spatial resolution, bandwidth, and the order of off-band images, etc.
In addition, in order to facilitate scientists to quickly acquire the off-band information, we synthesize the images and their off-band information into a JPEG picture, as shown in the Figure~\ref{Fig7}, in which the composite pseudo-color image of three off-bands is shown in the left panel, and the Line-of-sight (LOS) velocity map and monochromatic images are in the right respectively.
We can see that the composite pseudo-color image visually presents a well-aligned H$\alpha$ dataset. The fine-scale filamentary solar structures are sharp and clear without being smoothed or blurred by nonalignment.
It is worth mentioning that the LOS velocity is calculated by the center of gravity method from the information of off-band images (\citealt{Cai2017}), which may exist certain errors due to too few scanning wavebands. We list here just for the convenience of scientists' reference and quick browsing. To avoid misleading the data users, we wouldn't add the color bar to show the maximum and minimum value of the LOS velocity. If it needs to be used in scientific research, the algorithm for calculating LOS velocity of three off-band images would be further studied.
Furthermore, we make the sequence JPEG pictures into a movie for scientists to quickly browse the evolution of the observation target over time.

\section{Conclusion and Discussion}
\label{S:5}
The high-precision alignment between the winged images has always been an important problem restricting multi-band images joint research.
Based on the current situation of NVST H$\alpha$ imaging spectroscopic observation, we introduce a high-precision alignment algorithm for the off-band images based on the optical flow, and evaluate the alignment accuracy and criterion reliability under the different similarity (off-band) with the 2D spectral data of FISS/GST.
The research shows that the optical flow method can reach up to a high alignment accuracy in the near wing from line center, such as 0.004\,$''$ and 0.095\,$''$ in the off-band of 0.2\,\AA\, and 0.4\,\AA, respectively, which correspond the similarity of about 0.8 and 0.4. It can meet the alignment accuracy requirements of 0.1\,$''$ (within 1 pixel) under the similarity $\geq$\,0.3, which corresponds to the maximum offset wavelength of about 0.6\,\AA\,. Otherwise, it is necessary to increase more intermediate off-band images for transition alignment.
This result is helpful to select the appropriate offset wavelength for the routine three-bands observations of MHRIS/NVST.
After the fine alignment, we pack all off-band and line center images into three kinds of standard formats for the convenience of NVST data users.
In the paper, we mainly describe the application of the optical flow method in three off-band images. Actually, it is also suitable for the alignment of more off-bands images, and the alignment accuracy would be higher with the transition alignment of more off-band images.

In order to improve the accuracy of optical flow alignment algorithm as mentioned in the Section~\ref{S:2:2}, we make histogram statistics of the movement field on each pixel, and select a certain range near the peak of the histograms to calculate the reference position. Technically, the histogram distributions are different with the different similarity between aligned images, so the reference range should vary for different image compositions. However, we determine a fixed reference range by trial and error in this work, and will study an adaptive algorithm to select the reference range for further increasing the alignment accuracy in the future.
In addition, we use the off-band images composed by the 2D spectral data of FISS/GST to evaluate the alignment accuracy of algorithm. Those composed images are ideally strictly aligned. However, as the inevitable slight inaccuracy of the spectrometer system, the inclination or curvature of spectral lines always exists in the spectrum (\citealt{Cai2017}).
Although the curvature of spectral lines have been corrected in the post-processed spectrum (\citealp{Chae2004}), there is still a residual inclination in the spatial direction (the offset is about 10 pixels in the total wavelength range 10\,\AA\,). This residual inclination will systematically shift the composed off-band images in one spatial direction, and different off-band wavelengths correspond to different translation offsets. However, only the data with the wavelength within $\pm$\,1\,\AA\ of spectrum are used, and the bandwidth between the composed monochromatic images of two adjacent wavelengths is 0.1\,\AA, which corresponds to an inclination only about 0.1 pixel.
Therefore, to avoid the calculation error introduced by correcting residual inclination of spectral lines, the influence of the shifting is included in the values of RMSE, as shown in the Table~\ref{Tab1}. Therefore, the actual alignment accuracy should be higher than the result proposed in this paper.

\normalem
\begin{acknowledgements}
The authors are appreciated for all the help from the colleagues in the NVST team. We are also indebted to the GST teams for providing the FISS data.
This work is supported by the National Natural Science Foundation of China (NSFC) under grant numbers 11903081, 12073077, 11973088, 11873027, U2031140, 11833010, U1831210, Yunnan Natural Science Foundation of China (201901U070092), and the CAS ‘Light of West China’ Program under numbers Y9XB015 and Y9XB019.

\end{acknowledgements}

\bibliographystyle{raa}
\bibliography{bib}

\end{document}